# A Novel Indoor Mobile Localization System based on Optical Camera Communication

Md. Tanvir Hossan, Mostafa Zaman Chowdhury, Amirul Islam, and Yeong Min Jang
E-mail: mthossan@ieee.org

**Abstract**—Localizing smartphones in indoor environments offers excellent opportunities for e-commence. In this paper, we propose a localization technique for smartphones in indoor environments. This technique can calculate the coordinates of a smartphone using existing illumination infrastructure with light-emitting diodes (LEDs). The system can locate smartphones without further modification of the existing LED light infrastructure. Smartphones do not have fixed position and they may move frequently anywhere in an environment. Our algorithm uses multiple (i.e., more than two) LED lights simultaneously. The smartphone gets the LED-IDs from the LED lights whose are within the field of view (FOV) of the smartphone's camera. These LED-IDs contain the coordinate information (e.g., x- and y- coordinate) of the LED lights. Concurrently, the pixel area on the image sensor (IS) of projected image changes with the relative motion between the smartphone and each LED lights which allows the algorithm to calculate the distance from the smartphone to that LED. At the end of this paper, we present simulated results for predicting the next possible location of the smartphone using Kalman filter to minimize the time delay for coordinate calculation. These simulated results demonstrate that the position resolution can be maintained within 10 cm.

## 1. Introduction

Now-a-days, the number of mobile devices (i.e., smartphone) is dramatically increasing. These devices have an immense scope for commercial applications e.g., services, e-commerce, and e-banking. To move forward with consumer-facing commercial applications, location based services (LBS) for smartphones need to be improved. As the location of the mobile device is unpredictable; a reliable, dynamic, accurate, and situation-adaptive localization technique is required for LBS [1]. Moreover, this technique should be secure and interruption free. This LBS approaches will not increase consumer activity without a reliable localization system. The localization scheme design should be considered for indoor and outdoor environments. Indoor localization is most promising for e-commerce applications, as most of them are indoor based. The access density for smartphones is the highest in shopping malls, super markets, and transit stations (i.e., railway, bus, subway). Indoor environments are hubs for almost all web based business applications. Outdoor localization schemes are also promising and may be improved with insights from indoor localization solution.

Both industry and academic research institutes have recently shown interest in the issue of indoor smartphone localization, and various schemes have been proposed [2],[3]. The most common and widely used indoor localization system relies on the global positioning system (GPS) [4]. Moreover, this system has three particular limitations (e.g., poor GPS signal reception, loss of GPS signal, and limited localization accuracy) [5] especially in indoor environments. This system is not suitable for underground or indoor localization. Since the signal from the satellites to a receiver should be line-of-sight (LOS), and building, soil, water, trees or even poor weather conditions inhibits this signal. Time of flight (TOF) cameras are another possible candidate for solving the localization scheme [6]. Besides its advantages, this system is too expensive and sometimes requires complex scenarios for implementation which makes it inappropriate for unique approach. TOF camera also has some other drawbacks, it is only useful for detection and ranging purposes and does not facilitate the communication purpose [7]. Received signal strength indication (RSSI), time of arrival (TOA), time difference of arrival (TDOA), angle of arrival (AOA) are physical parameters of radio signal that can be used for localization with special distributed monitors [8],[9]. Other approaches used for indoor localization include computer vision (CV) and artificial intelligence (AI) [10]. None of these approaches are ideal for indoor localization. For example, RSSI depends on environmental conditions. It is affected by shadowing, path loss, signal fading, and interference from neighboring cells [11]. Therefore, errors can be included in the calculated value. On the contrary, narrowband signals, reduced data transmission rate, and lower location precision can be inhibited in TDOA parameter [12]. Other approaches like combining RSSI and AI cannot mitigate the challenges in indoor localization because subjective and objective data from AI have an impact on new input data [13]. On the basis of system performance improvement, it may feedback to the input of the system. Concurrently, it will cause a huge impact on new data if input variables are varying fast. Another approach, photogrammetry, measures object location from the photographs. This is a very simple way to generate a map from sequentially taken photographs [14].

To deal with the existing challenges for indoor localization system, optical frequency band in electromagnetic spectrum is a novel candidate that avoids the problems typical with radio frequency (RF). The communication technology where these optical frequency is using in known as optical wireless communication (OWC). A promising sub-system of OWC is optical camera communication (OCC), which multiple LEDs are used as transmitters and a camera or image sensor (IS) is used as receiver [15]. In an indoor environment, the communication channel for OCC is un-interrupted, signal-to-noise (SNR) is high, security is ensured by LOS communication, simple signal processing is confirmed, and high speed communication are possible [16]. OCC has multiple-input-multiple-output (MIMO) functionality [17] which is ideal for analyzing many objects at the same time. Therefore, OCC is a good technique for indoor localization systems. Moreover, OCC is not only useful for localization but also for communication. Since 2011, the IEEE has formed a new working group to finalize the standardization specification (i.e., IEEE 802.15.7) [18]. The development of the standardization specification for OCC will be finalized by middle of 2018.

In the proposed scheme, smartphones are located through signal from LED light using OCC and photogrammetry [19]. Using the OCC techniques, smartphone cameras receive the LED-ID signals from each LED light (or fixture). This LED-ID resemble coordinate (x, y) of the LED lights. The size of the image of the LED fixture on the image sensor of the camera changes with the relative distance between the smartphone and LED fixture. This distance is calculated using photogrammetry which helps to find the remain coordinate of the smartphone. To optimize the resolution, Kalman filter is applying to accurately predict the next possible locations of the smartphone.

The rest of paper is organized as follows. In Section 2, we survey several literatures on indoor localization system. Our proposed scheme is illuminated in Section 3. In Section 4, the channel modeling and communication are explained. The processes for distance calculation and localization is stated in Section 5. In Section 6, the performance of our proposed system using Kalman filter is evaluated. Finally, we summarize our work and future research direction in Section 7.

## 2. Related Works

In this section, several indoor localization schemes are summarized. OWC based indoor localization can be classified in three ways: triangulation, fingerprinting, and proximity-based approaches. A geometric position is required in triangulation. Fingerprinting is scene analysis approach and proximity-based approach is grid method. In [20], a vision-based positioning and navigation is introduced with the help of a camera and newly defined 3D map in an indoor environment. The authors tried the possible ways to improve the accuracy and system reliability through several experiments. They mainly focused on different applications e.g., mobile robot navigation, transportation, visitor guiding, security, and emergency services. In the same manner, authors in [21], propose a vision-based system using a camera as the main sensor for object detection, tracking, and localization for distributed sensing, communication, and parallel computing. They used camera nodes as inputs of sensor fusion techniques to reduce computer system encumbrances such as Extended Kalman Filter (EKF) and Maximum Likelihood (ML). They also test investigation for moving objects. The drawback of their proposed system is the lacking of automatic determination of sensor location. For an indoor environment, a vision-based navigation system using augmented reality was proposed in [22]. They recognize a location automatically using image sequences which is taken in the indoor environment and then realized augmented reality by flawlessly superimposing the user's view with detailed location information. They use wearable mobile PC with a camera for taking image sequences and it transmits the images to remote PCs for future processing. Using their proposed system, the average location recognition success rate was found around 89%. Moreover, its performance will deteriorate with harsh environmental scenarios. In [23], neural network and OCC based indoor positioning system is proposed. They estimate camera position for trained and untrained environments. The error for estimated camera position is less than 10 mm, which can increase up to 200 mm. However, the error depends upon the camera position. In [24], authors demonstrated a system for positioning and orientation of overlying the location information on camera phones in an indoor environment. They find a location from images with the help of existing standard hardware. They process less data with in very short time to generate accurate data. The limitations of their proposed system in handling the whole environment is that they must improve the feature detection and match while maintaining low latency. Their proposed system is complex and very costly for navigation on a standard camera phone. In [25], authors propose radiometry and camera geometry for OCC camera model. They consider a camera model for the indoor environment at 50 cm separation for LEDs and the distance between LEDs is 200 cm.

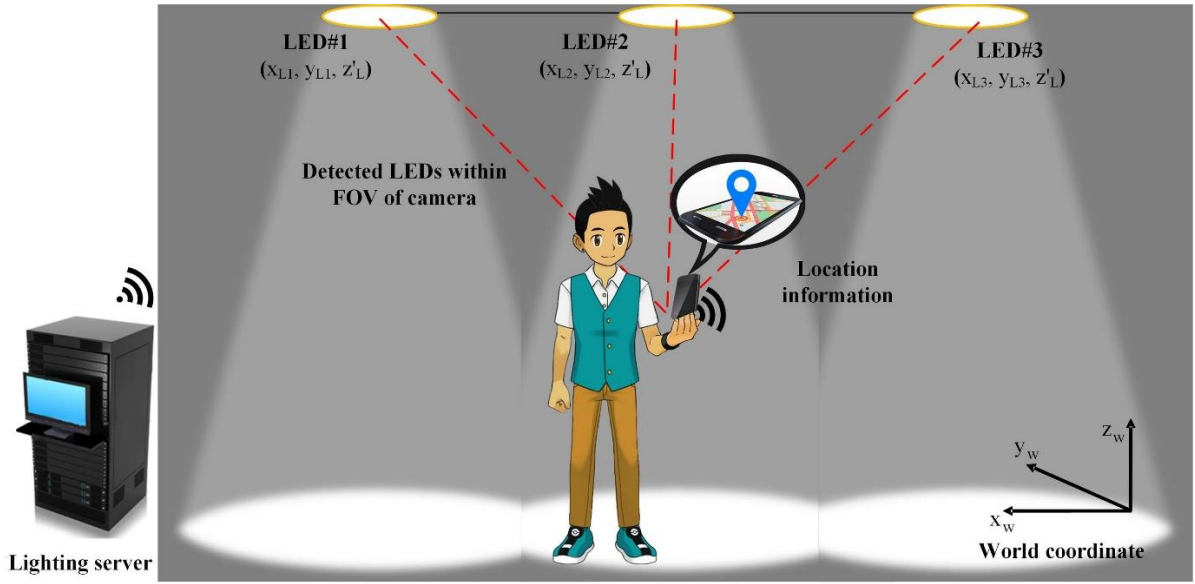

FIGURE 1: Overall architecture of proposed indoor localization scheme.

## 3. Overall Architecture

We propose a localization scheme in which smartphones can be located in an indoor environment with the help of LED lights and the smartphone's camera. In our proposed scheme, we consider several factors to identify smartphone locations in indoor environments. As in Figure 1, all LED lights fixture are attached to the ceiling, the distance between celling and floor is constant for particular indoor environment, the camera of smartphone must be under the illumination of the LED light fixture, and there must be at least three LED lights in the field of view (FOV) of the camera, while the smartphone maintains continuous communication with the lighting server. A lighting server provides API-based access to reproducible, web-based visualizations. The FOV of a camera is the solid angle through which the IS can sense electromagnetic radiation such as visible light. The system performance improves if the number of LED lights within the camera FOV is increased and our algorithm requires at least three light fixtures to deliver accurate location measurements. For our system, the LED lights fixtures coordinates are parallel to the world coordinate system. The vertical distance between ceiling to floor (i.e., z-coordinate) is equal for all LED lights where the z-coordinate of light fixture is in the inverse direction to the world z-coordinates. However, the camera coordinates of the smartphone change frequently with respect to the LED light fixture coordinates.

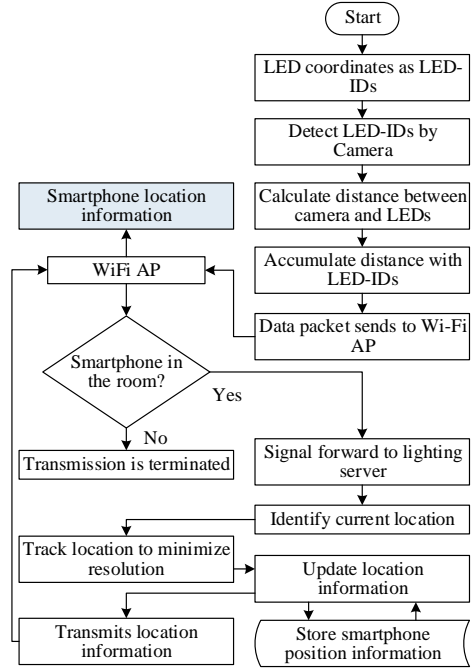

FIGURE 2: Flow chart for proposed scheme.

As shown in Figure 2, each LED light broadcast its own coordinate information (i.e., x, y coordinate) as a modulated LED-ID signal. The z-coordinate is same for each and every LED light in a certain indoor environment. Therefore, the term of z-coordinate in the LED-ID is ignored to reduce the complexity and data packet size. These LED-IDs are transmitted as a modulated optical signal. After receiving the signal from the LED lights, smartphones process the data in two different ways. They identify LED-IDs from received signals and measure the distance of the LED light with the corresponding LED-ID using photogrammetry. This distance is measured by calculating the size of the light fixture and number of pixels of the corresponding LED light on the IS. The geometric image size of the LED on the IS varies with the distance between the light source and camera. If the LED is located far away from the camera, the size of the image is smaller on the IS and comparatively a large image will generate for the LED light which is near to the camera.

LED-IDs with the corresponding distance calculation are sent to the lighting server by the smartphone via a wireless fidelity (Wi-Fi) access point (AP). The coordinate information for each LED lights is stored in the lighting server. After receiving a signal from the smartphone, the lighting server matches the LED-ID signal with its stored location information. Then the algorithm can map the location of the smartphone.

Meanwhile, the location of the smartphone may change during this processing time. Therefore, the lighting server uses Kalman filter tracking algorithm which predicts the next possible location from the current location of the smartphone. This location information sends to the smartphone. Additionally, the LED light's projected image on the IS changes as the smartphone moves. Therefore, the placement interval of the LED lights should be constant distance to ensure that at least three images of LED lights are available from any location of the room.

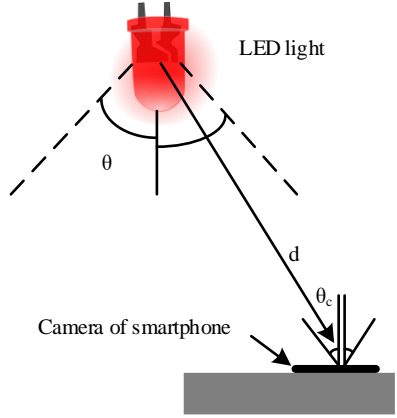

FIGURE 3: Geometric model for LOS communication.

## 4. Channel Modeling and Communication

*4.1. Propagation model of light from LEDs.* The radiation pattern from LED lights is affected by the roughness of the chip faces, and geometry of the encapsulating lens. Several models are used to describe the direction of light strength from a light source (i.e., LED). A popular approach is Monte Carlo ray tracing [26]. In a Gaussian (or cosine) power distribution [27], the ray of light is diffusely reflected or refracted. The final radiation pattern of light should appear linear super which are angularly shifted in function of the angle of incidence of every traced ray due to diffusely reflection or refraction characteristic shown in Figure 3.

The energy flux per solid angle known as luminous intensity and transmitted optical power are the two basic properties of a light source such as LED light. The luminous intensity is given as

$$I(\phi) = I(0)\cos^m(\phi) \qquad (1)$$

where $\phi$ is the luminous flux and $I(0)$ is the center luminous intensity of an LED. This luminous flux can be defined as integration between minimum wavelength, $\lambda_1$ to maximum wavelength, $\lambda_1$ of working optical spectrum

$$\phi = K_m \int_{\lambda_1}^{\lambda_2} V(\lambda)\phi_e(\lambda)d\lambda \qquad (2)$$

where $V(\lambda)$ is the standard luminosity curve and $K_m$ is the maximum spectral efficacy for vision. The integral of the energy flux $\phi_e$ in all directions is the transmitted optical power $P_t$, given as

$$P_t = \int_0^{2\pi} \phi_e d\theta d\lambda \qquad (3)$$

Figure 4(a) shows the LED light propagation direction and Figure 4(b) shows the strength of the radiation from a single light source. The power level is maximum 40 dBm at the center of the light source which is represented as yellow in Figure 4(a) and deteriorates to -80 dBm represents as violet; from the center to the edge of the sphere. Joining average power strengths of light at certain x- and y-coordinate is forming ellipse shape in Figure 4(b). Therefore, weak signal from neighboring light sources cause signal interference. This problem can be mitigated by removing the background noise with a full control on camera shutter speed.

*4.2. LED-ID for optical camera communication.* Each LED light has a fixed location and the coordinate information for each a single LED light is different from the other LED lights in the same room. The coordinates of the LED light are parallel to the world coordinates. These coordinates are analogues to the LED-ID. Every LED light transmits its own ID to the camera which can be declared as a digital tag. For this purpose, the LED acts as a transmitter and camera as a receiver.

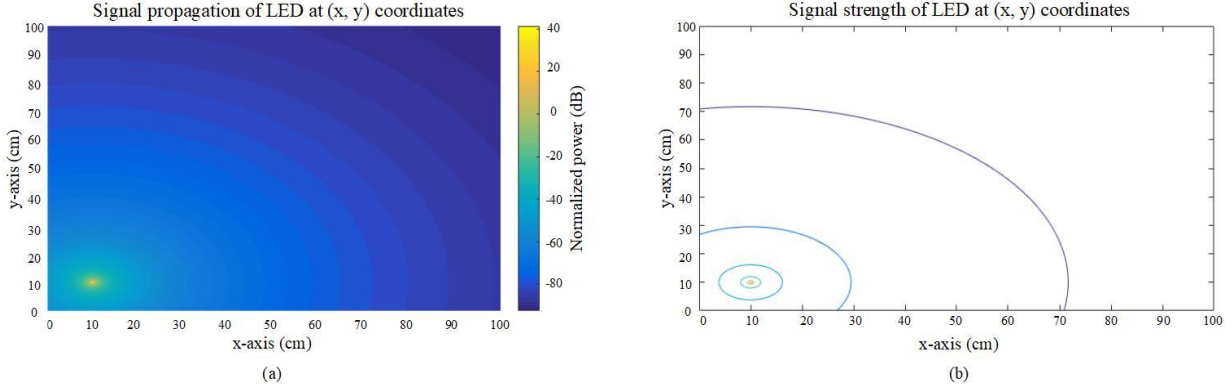

FIGURE 4: LED light (a) propagation and (b) average strength at (x, y) coordinates.

The data from the LED light is sent as a modulating signal by varying the intensity of the light using LED driver integrated circuit (IC). This driver controls the light intensity by dimming the LED through a variety of methods. Data from the LED light can be encoded in the phase of LED light signal and phase can be changed by turning the LED light on and off. However, turning the LED light on is not always possible after turning the lights off fully. Therefore, it is recommended to dim the light at minimum intensity. Here, this modulation is known as IM/DD (Intensity Modulated/Direct Detection) modulation [28]. The available modulation technique to transmit signal from LED light sources can be classified as

- On-off keying (OOK) [29]: The two logic signals in a digital transmission '1' and '0' are represented as high and zero voltage at the transmitter end. This is achieved by applying flickering illumination of the LED light to represents the on-off state of the transmitter.
- Pulse width modulation (PWM) [30]: The modulated signal from the LED light is transmitted in the form of a square wave. The desire level of the pulse is obtained by adjusting the LED light dimming.
- Pulse-position modulation (PPM) [31]: Light from the LED encoded message is transmitted through a single pulse in one time shifts.
- Orthogonal frequency division multiplexing (OFDM) [32]: Data is sent as parallel sub-streams of modulated data using multiple orthogonal subcarriers in a channel.
- Frequency-shift keying (FSK): A modulated digital signal can be carried by the instantaneous frequency shifting with a constant amplitude.
- Phase-shift keying (PSK) [33]: A digital signal can be carried by the instantaneous phase shifting of the baseband signal.

For high speed data transmission channels, OFDM is used where the possibility of multipath fading and inter-symbol interference is high [34]. Despite the enormous advantage afforded by OFDM, we do not implement it, because high-speed data transmission is not required for most indoor localization applications. In these cases, OOK is a better choice for data transmission from LEDs. Figure 5 shows how the after encoding and modulating the light, the LED light fixture transmits data to the camera of the smartphone. Our system decodes LED coordinates, after demodulating and decoding the received signal from the image processor.

*4.3. Channel modeling for OCC.* The pixel $E_b/N_0$ of the model of IS [35] can be calculated as

$$Pixel \frac{E_b}{N_0} = \frac{E[\rho^2]}{E[n^2]} \approx \frac{s^2 \Delta}{\alpha s \Delta + \beta} \qquad (4)$$

where $E_b$ represents the energy-per-bit, $N_0$ represents the Spectral-Noise-Density, $\rho$ is the unit pixel value, $s$ is the amplitude of the signal, $\Delta$ is the camera exposure duration as a ratio of the signal cycle, $n$ is a noise term, and $\alpha, \beta$ are model fitting parameters for system noise.

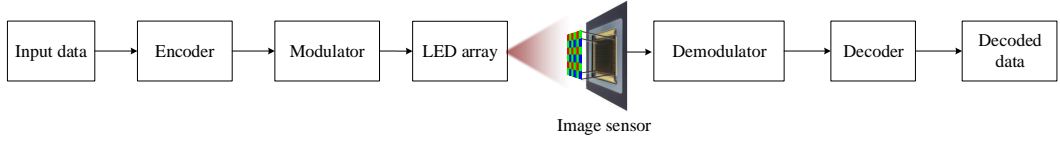

FIGURE 5: Basic architecture of optical camera communication.

Considering the distortion in the channel, the signal-to-noise-plus-interference ratio (SNIR) can be determined as

$$SNIR = \frac{\sigma P_{avg}^2}{(1-\sigma)P_{avg}^2 + \sigma_n^2} \qquad (5)$$

where $P_{avg}$ is the average pixel intensity transmitted from the LED light, $\sigma$ is the average distortion factor which value lies with 1 to 0 i.e., $0 \leq \sigma \leq 1$. $\sigma = 1$ indicates minimum signal lost with the LED light is focusing directly on to the camera lens and $\sigma = 0$ indicates that no image pixel is generated on the image sensor of the camera.

Since the additive white Gaussian noise (AWGN) characteristic on the camera channel [36], the channel capacity of the space time modulation can be expressed by Shannon capacity formula as

$$C = F_{fps} W_s \log_2(1 + SNIR) \qquad (6)$$

where $F_{fps}$ is the frame rate of the smartphone camera, $W_s$ is the spatial bandwidth, which represents how much information is carried by the pixels in each image frame. The spatial bandwidth is equivalent to the number of orthogonal or parallel channels in a MIMO system.

The bit error rate (BER) which depends on the SNIR and the modulation scheme measures the impact of the channel. The noise sources that affect the transmission of the light signal from the LEDs are inter-symbol interference, background and transmitter LED shot noise, and thermal noise.

Contemporary smartphones use complementary metal oxide semiconductor (CMOS) based IS where shutter type is rolling shutter. With this shutter technique, light intensities on the IS are captured row by row and the whole image is composed of different pixel array. Therefore, the exposed time delay between pixel array lines records the changing state of illumination of the LED light as a group of pixels in one image. The optical channel DC gain $H(0)$ models the channel characteristic from LED lights to the camera which can be determined [37] as

$$H(0) = \begin{cases} \frac{(m+1)A_{cam}}{2\pi d^2} \cos^m(\phi) T_s(\theta) \cos(\theta), & 0 \leq \theta \leq \theta_c \\ 0, & \theta \geq \theta_c \end{cases} \qquad (7)$$

where $m$ is the order of Lambertian emission, $A_{cam}$ is the area on IS, $d$ is the distance between an LED and IS, $T_s(\theta)$ is the optical filter coefficient for signal transmission, $\theta$ is the angle of incidence, and $\theta_c$ the is camera FOV semi-angle. Here, $m$ can be define as

$$m = -\frac{\ln 2}{\ln(\cos_{\theta_c/2})} \qquad (8)$$

If $N$ is independent of signal characteristic and is used to calculate the channel output $v$ for each transmitter $u$ for $n$ number of LED-ID signal as

$$v = R \sum_{i=1}^{n} h_i x_i + N \qquad (9)$$

where $u = [u_1 \ u_2 \ ... \ u_n]^T$ and $R$ is the camera responsivity.

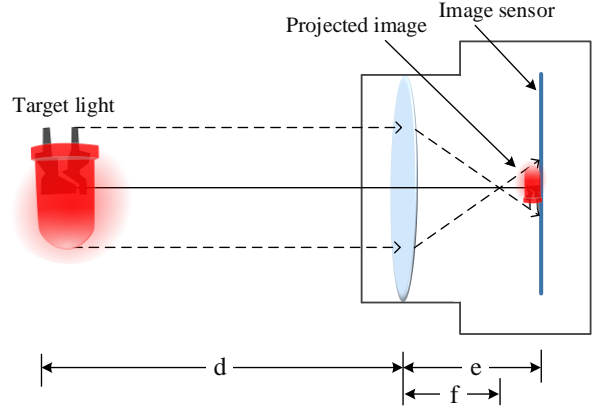

FIGURE 6: Illustration of projected image on the image sensor.

The average optical power of the IS of the camera can be calculated as

$$P_r = \frac{I(0)\cos^m(\phi)\cos(\psi)}{d^2} \tag{10}$$

## 5. Distance Calculation and Localization

*5.1 Distance calculation between LED light and camera.* The fundamental operation of a camera is diagrammed in Figure 6 where an image of a target LED light is projected on the IS of a camera. Light from the target LED passes through the camera lens and is concentrated on the IS plate. The projected image on the IS plate is an inverted image of LED light fixture.

Consider that $f$ is the focal length of the camera, $d$ is the distance from the camera lens to the target LED light, and $e$ is the distance from the focal length to the projected image on the IS. Therefore, we can write

$$\frac{e}{d} = \frac{f}{d-f} \tag{11}$$

The magnification of the lens is the ratio of the projected image size to the geometric size of LED light. If the camera projects a square image on the IS where the height and width of the target LED and the projected image are ($a$, $b$) and ($a_i$, $b_i$), respectively then the lens magnification can be expressed as

$$M = \frac{a_i}{a} = \frac{b_i}{b} = \frac{e}{d} = \frac{f}{d-f} \tag{12}$$

If we ignore the loss in the optical channel then $d \gg f$. More precisely $d - f \approx d$. By combining (11) with (12), we get

$$a_i b_i = M^2 ab \tag{13}$$

The number of pixels of image is the ratio of the projected image size on the IS to the unit pixel area of the same sensor. If the number of pixels is $\eta_i$ on the IS, $\rho^2$ is the unit pixel area of IS, and $S$ is the area of the target LED light source, then we obtain

$$\eta_i = \frac{f^2 S}{\rho^2 d^2} \tag{14}$$

Different shapes of light fixture are available in the market. However, rectangular/square and circle are typical light fixture shapes. For circular shape LED light fixture, if $a$ is the radius of the circle then the area of the light source will become, $S = \pi a^2$. On the contrary, if $a$ and $b$ are the width and length of a rectangular

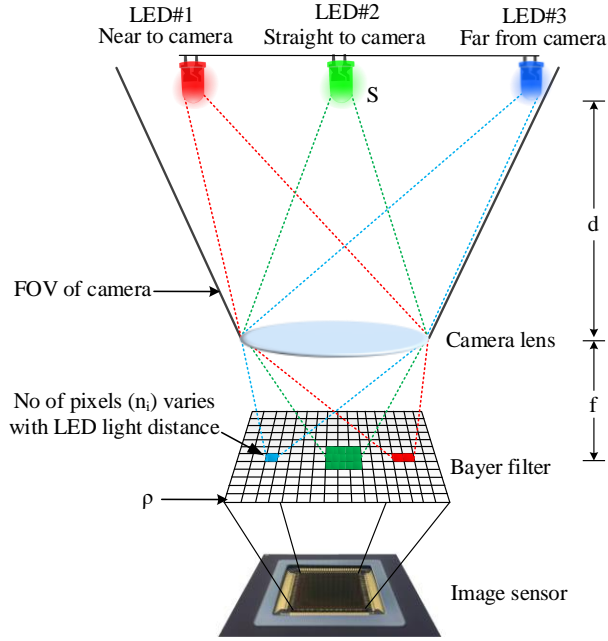

FIGURE 7: Calculating distance from LED light to camera by the image area on the IS.

light sources respectively, then the area of the light source will be, $S = ab$.

If we only consider the LED lights within the smartphone FOV, the distances from the LED light to the smartphone camera are different for each and every lights. The projected image from LED light facing the camera straight-on, is larger than the image of an LED light located at an angle to the same camera. These distances can be determined by (14) which must be modified due to the relative motion between the camera and LED light fixtures. For each camera, the focal length is $f$, and the unite pixel area of IS are fixed. On the contrary, if we know the surface area of LED light fixture. We can write (14) as

$$d = \tau \sqrt{\frac{1}{\eta_i}} \qquad (15)$$

where $\tau = f \frac{\sqrt{S}}{\rho}$ is constant for each camera and LED light.

Figure 7 shows how the image areas are varying with the distance of the LED lights from the camera. From (15), the distance from IS to the LED light fixture is inversely proportional to the square root of the image area on the IS.

*5.2. Scenario of image area on the IS for dynamic of camera.* The smartphone is the only moving device with respect to the LED lights in indoor environments. The FOV of the camera shifts with changes in the smartphone's location. Therefore, the images of LED lights on the IS of camera also change. Additionally, the lighting infrastructure must be designed in a way that keeps at least three LED lights within the FOV of each camera. Figure 8 describes a scenario in which the LED lights detected by the camera vary from 4 to 5 due to the movement of the smartphone from location 1 to location 5.

Square blocks, black dots, and circles represent LED lights, camera, and the FOV of the camera, respectively. For location 1 of the smartphone, five LED lights images are projected on IS. The number of projected image varies from 5 to 4 when the location of the smartphone changes from location 1 to location 5. If more than three LED lights images are projected on the IS, location of the smartphone can be calculated more accurately.

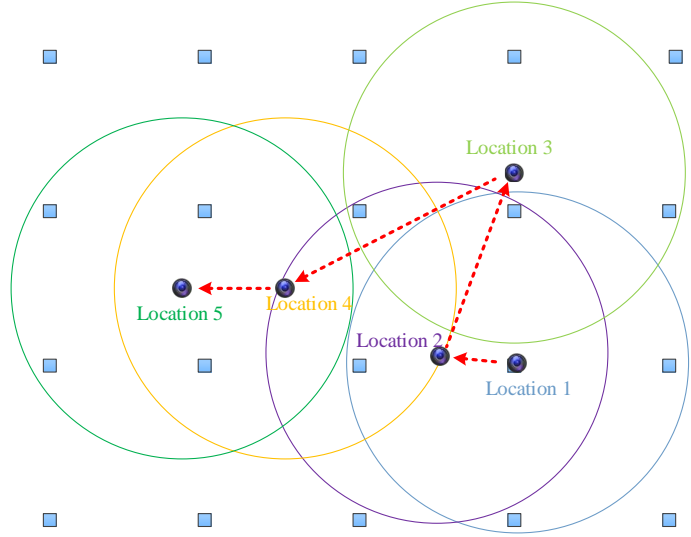

FIGURE 8: Change of considered LED lights within the FOV of camera with the change of smartphone.

Next, we consider another scenario in which the smartphone moves from location 1 to location 2, shown in Figure 9 (b). Due to the change of smartphone location, the image of the LED lights on the IS also changes. We consider three possible location of LED lights (e.g., red, green and blue), all attached to the celling. For location 1 of the smartphone in Figure 9 (c), the blue LED light is relatively close to the camera. This means that the angle is smaller from camera to the blue LED light. This angle is large for the red LED light compared with the blue LED light. On the contrary, this angel is nearly zero for the green LED light. These distances are calculated using (15).

We describe the straight-line distances of red, green and blue LED lights from lens of the camera are $d_{r1}$, $d_{g1}$, $d_{b1}$, respectively. Furthermore, $\eta_{ir1}$, $\eta_{ig1}$, $\eta_{ib1}$ are the number of pixels on the IS for red, green, and blue light fixtures, respectively at location 1. Therefore, (15) can be written for each LED lights as

$$d_{r1} \propto \frac{1}{\sqrt{\eta_{ir1}}} \qquad (16)$$

$$d_{g1} \propto \frac{1}{\sqrt{\eta_{ig1}}} \qquad (17)$$

$$d_{b1} \propto \frac{1}{\sqrt{\eta_{ib1}}} \qquad (18)$$

From Figure 9 (c), the size of image on the IS is larger for the green LED light. This size gradually decreases from the blue to red LED lights. Therefore, comparing one particular image with the other image on IS can be mathematically represented as $\eta_{ig1} > \eta_{ib1} > \eta_{ir1}$. As we know from (15), the number of pixels projected on the IS for a certain object depends only on the distance between the camera and the object when other factors remains constant. The mathematical expression for each and every image on the IS and the (16)-(18) gives a conclusion: $d_{r1} > d_{b1} > d_{g1}$.

At location 2 in Figure 9 (a), the angular distance between the camera and the blue LED light fixture increases compared with the red LED light. Therefore, The image size increases accordingly. In contrast, the image size of the green LED light is almost identical due to the small shift in angle.

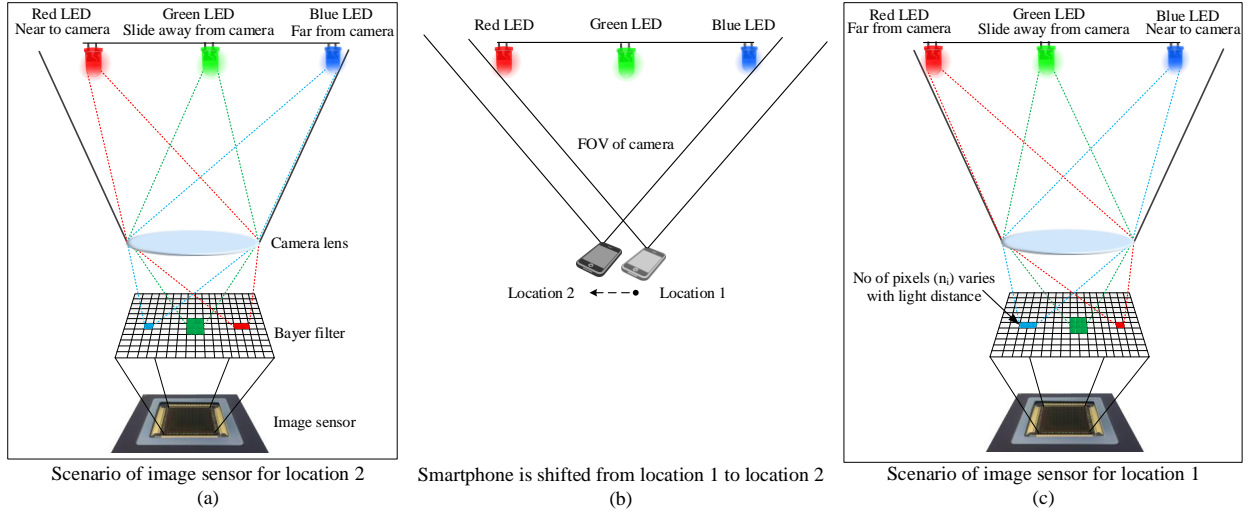

| Scenario of image sensor for location 2 | Smartphone is shifted from location 1 to location 2 | Scenario of image sensor for location 1 |
|---|---|---|
| (a) | (b) | (c) |

FIGURE 9: Scenarios of pixel size on IS for the movement of smartphone from location 1 to location 2; (a) position 2; (b) shift of FOV of the camera; (c) location 1.

For location 2, the distance from LEDs lights to the camera lens are notated as $d_{r2}$, $d_{g2}$, $d_{b2}$ for red, green and blue LEDs, respectively. Moreover, the numbers of pixels on the IS are labeled as $\eta_{ir2}, \eta_{ig2}, \eta_{ib2}$. Therefore, if we apply (15) for three LED light images, we will get

$$d_{r2} \propto \frac{1}{\sqrt{\eta_{ir2}}} \tag{19}$$

$$d_{g2} \propto \frac{1}{\sqrt{\eta_{ig2}}} \tag{20}$$

$$d_{b2} \propto \frac{1}{\sqrt{\eta_{ib2}}} \tag{21}$$

From Figure 9 (a), the general expression for the number of image pixel is $\eta_{ig2} > \eta_{ir2} > \eta_{ib2}$ on the IS. From (19)-(21), the distance can be explained as: $d_{b1} > d_{r1} > d_{g1}$.

*5.3. Uploading information to the lighting server*. Thea smartphone begins to send distance information with the corresponding LED-ID to the lighting server via Wi-Fi AP. This information is sent as a packet with two slots where the first slot has the coordinate information of the LED light and second slot has its distance information. The information of LED light coordinates is already stored in the lighting server. After receiving a signal from the smartphone, the lighting server generates a virtual map of LED lights from extracting information from the packet. With the mathematical model of trilateration (or multilateration for more than three LED lights), the lighting server calculates the location of the smartphone.

*5.4. Computing smartphone location.* The beam of LED light propagates from ceiling to floor and illumination spreads 360° from the center of a light source. At a specific distance from the LED light, the intensity of light illumination is equal. If the smartphone is located at any of these locations; the LED image will be the same size on the. Therefore, the chance of false location identification is higher for a single LED light source. Figure 10(a) shows the original location of the camera with the probable false location of the camera (in a faded image). This false location information creates errors during location mapping in the lighting server.

Consequently, to mitigate location-estimation errors when only a single LED light is visible, we consider another LED light as a reference for the first LED light. From Figure 10(b), we narrow down the location to the torus formed by the intersection of the two spheres. Though more LED lights are available in this scenario, the duplication still occurs. Hence, a third LED light is necessary to deliver accurate location measurements.

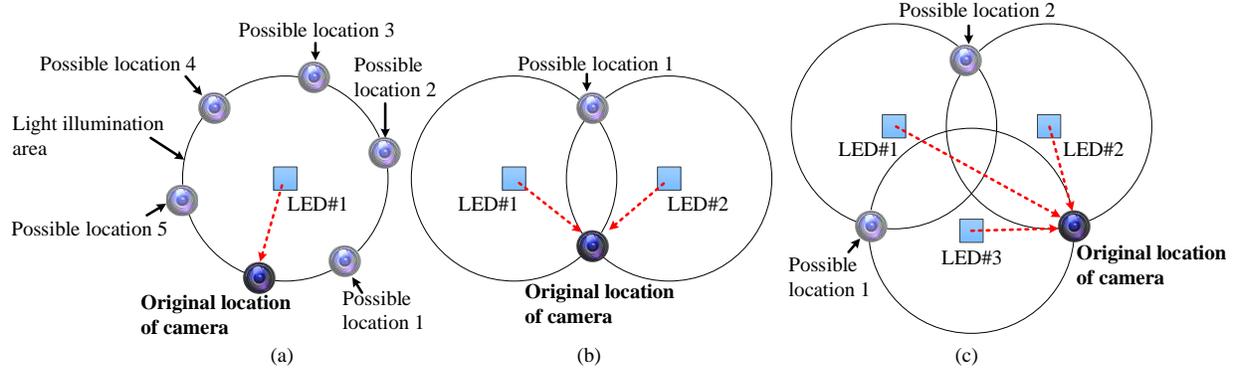

FIGURE 10: Narrowing down localization estimation error using (a) single LED light (b) two LED lights, and (c) three LED lights.

With a third LED light we can narrow the location of the smartphone to one possible location. In Figure 10(c), three circles centered on each of the landmarks overlap at three different locations where the radius of each circle is equidistant from each landmark. Therefore, two other locations of the smartphone along with the original are still possible even with three landmarks. Moreover, other location information may not arise any confusion for smartphone position estimation. It is possible to estimate the smartphone's location accurately by comparing information from two LED lights with information from a third light.

The method to determine the location of smartphone using three fixed reference points (or LED lights) is known as trilateration [38] or more than three points (which is known as multilateration). For trilateration, the measuring platform is simultaneous with three relevant nonlinear equations. The reference LED lights can be situated either in a triangle or in a straight line from each other.

If $P_i = (x_{Lj}, y_{Lj}, z_L')$ is the coordinates of any LED light under the celling where $j = 1, 2, 3$ and the coordinate information of the smartphone camera are $P_c = (x, y, z)$, then the distance from the camera to the LED light can be represented with following equation

$$d_j^2 = (x - x_{Lj})^2 + (y - y_{Lj})^2 + (z - z_L')^2 \tag{22}$$

A matrix can be generated with the above equations for $i = 1, 2, 3$ as

$$\begin{bmatrix} 1 & -2x_{L1} & -2y_{L1} & -2z_L' \\ 1 & -2x_{L2} & -2y_2 & -2z_L' \\ 1 & -2x_{L3} & -2y_{L3} & -2z_L' \end{bmatrix} \begin{bmatrix} x^2 + y^2 + z^2 \\ x \\ y \\ z \end{bmatrix} = \begin{bmatrix} d_1^2 - x_{L1}^2 - y_{L1}^2 - z_L'^2 \\ d_2^2 - x_{L2}^2 - y_{L2}^2 - z_L'^2 \\ d_3^2 - x_{L3}^2 - y_{L3}^2 - z_L'^2 \end{bmatrix} \tag{23}$$

The matrix equation can be replaced as

$$Z_0 x = q_0 \tag{24}$$

Two different cases can occur when solving the trilateration problem for locating the smartphone. The LED lights can be distributed randomly as in Figure 11(a) or aligned in a straight line as shown in Figure 11(b).

To identify the location of a smartphone from the reference LED lights located at the vertexes of a triangle, the general solution of (24) can be expressed as

$$x = x_p + tx_h \tag{25}$$

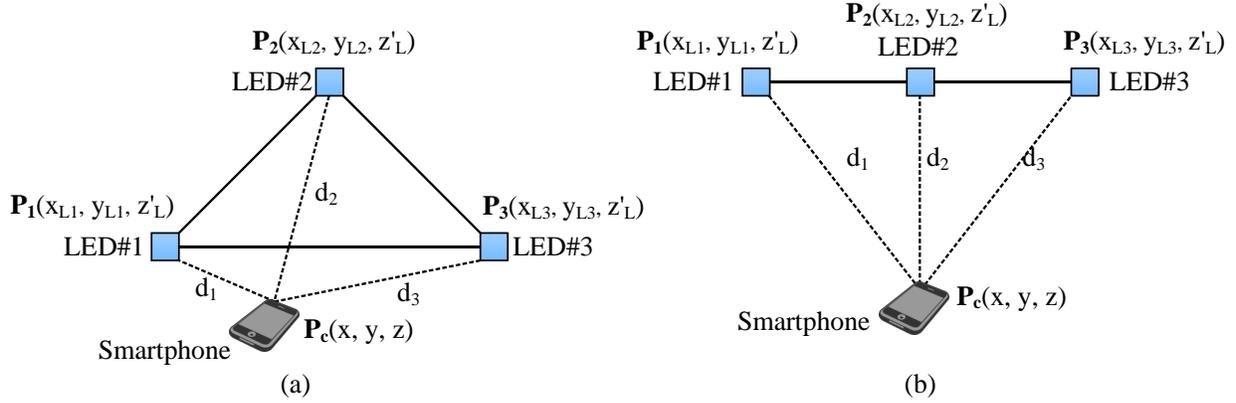

FIGURE 11: Determination of camera location from reference LED lights those are located at (a) the vertexes of triangle and (b) different points on a straight line.

where $x_p$ is denoted as the particular solution $t$ is the real parameter. If $Z_0 x = 0$ is a homogeneous system then $x_h$ is its solution.

The matrix $Z_0$ is written as pseudoinverse matrix format to determine the solution for $x_p$. On the other hand, the value of $t$ can be evaluated using the expression of $x_p = \begin{bmatrix} x_{p0}, & x_{p1}, & x_{p2}, & x_{p3} \end{bmatrix}^T$, $x_h = \begin{bmatrix} x_{h0}, & x_{h1}, & x_{h2}, & x_{h3} \end{bmatrix}^T$, and $x = \begin{bmatrix} x_0, & x_1, & x_2, & x_3 \end{bmatrix}^T$.

Following solution can be generated after solving (25) as

$$x_1 = x_p + t_1 x_h \tag{26}$$

$$x_2 = x_p + t_2 x_h \tag{27}$$

To identify the location of a smartphone from reference LED lights located in a straight line, the general solution of (24) is expressed as

$$x = x_p + t x_{h1} + \kappa x_{h2} \tag{28}$$

where homogeneous system $Z_0 x = 0$; $x_{h1}$ and $x_{h2}$ are two solutions with real parameters $\kappa$.

The mathematical expression of (24) is different for the case with more than three LED lights. The solution can be found by solving multilateration problem. The relevant equation can be expressed as follows:

$$Zx = q \tag{29}$$

On the base of the least squares methods, the solution of (29) can be found as

$$\hat{x} = (Z^T Z)^{-1} Z^T q \tag{30}$$

In Figure 12, three LED lights are located at three points (i.e., $P_1$, $P_2$ and $P_3$) in a two-dimensional plane. Their illumination spheres intersect at two points (i.e., $P_{C1}$ and $P_{C2}$) that are possible locations for the smartphone camera. The lighting server chooses between the multiple possible location of the smartphone using trilateration.

Figure 13 shows that systematic estimation error is generated when the lighting server estimates the position of the smartphone. The error is minimum for the horizontal bias (in Figure 13 (a)) of the indoor environment and is much higher for the vertical bias (in Figure 13(b)). Therefore, system performance is much degraded when measuring vertical position.

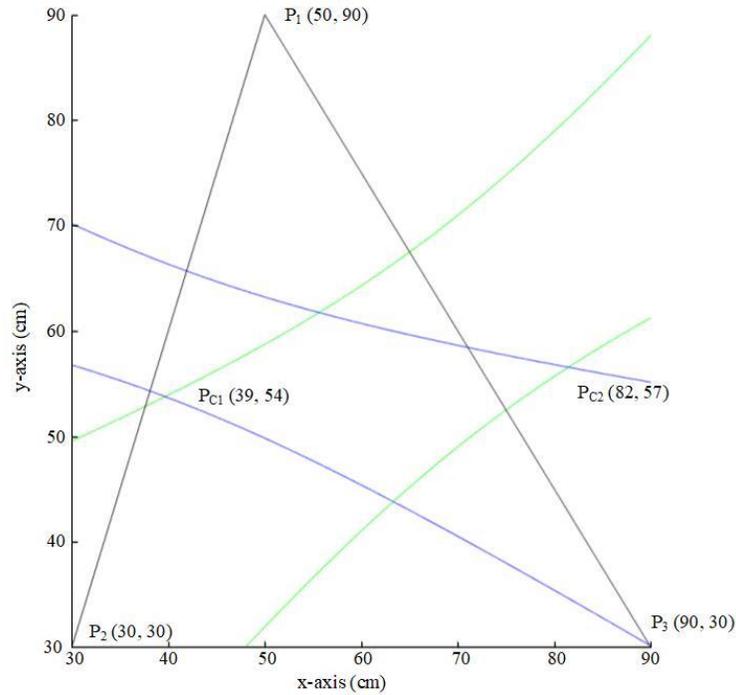

FIGURE 12: Trilateration using three LEDs in 2D plane.

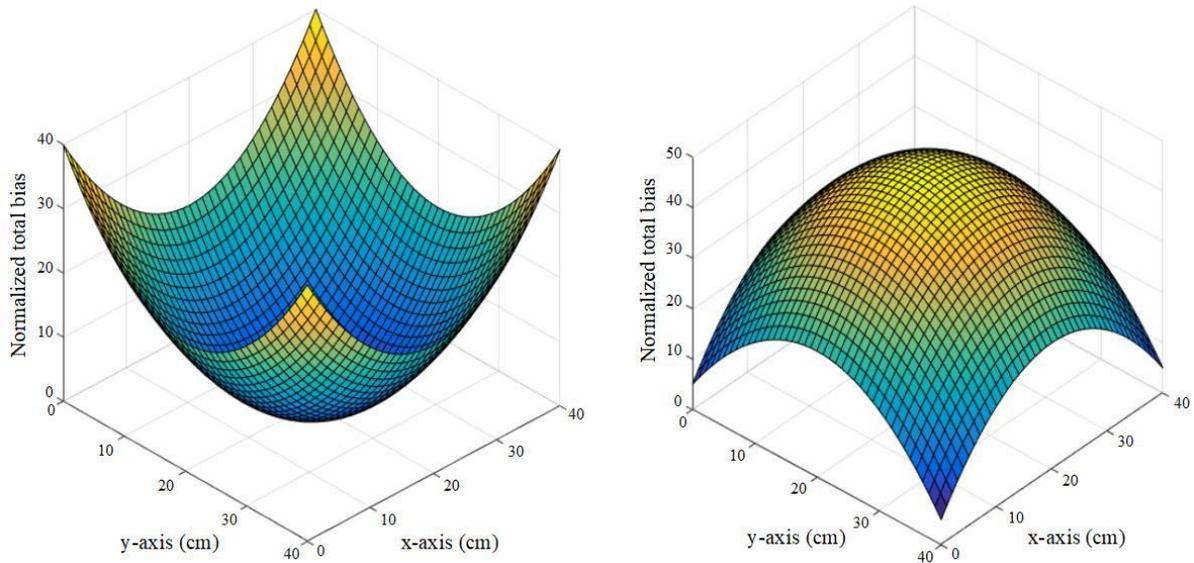

FIGURE 13: Systematic errors in (a) horizontal (b) vertical direction.

Around a cluster of three LED lights, the possible locations of the smartphone are shown in Figure 15. Dotted lines represent the optical links between the camera and the LED lights and solid lines represent the fixed distances between the LED lights. In almost all cases the distance from the smartphone to each LED light is different. In some cases, (Figure 14(b), (c), (d), (g)) the distance to two LED lights is equal compared with the distance from the other LED light. Additionally, there are few cases (Figure 14(e), (f), (h)) where all three distances are different from each other. Concurrently, there is only one case (Figure 14(a)) where camera is equidistant from all LED lights. The algorithm can locate smartphones at these locations without error.

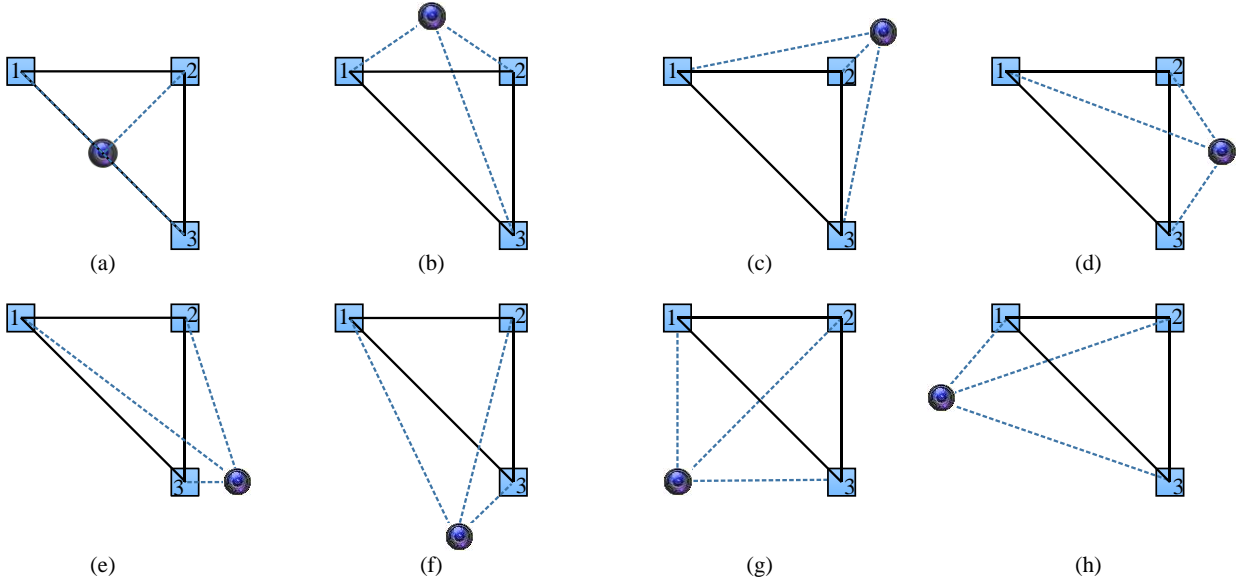

(a) (b) (c) (d)

(e) (f) (g) (h)

FIGURE 14: Relative distance from LED lights to camera of smartphone; (b), (c), (d), (g) two equal distances; (e), (f), (h) different distance; (a) equal distance.

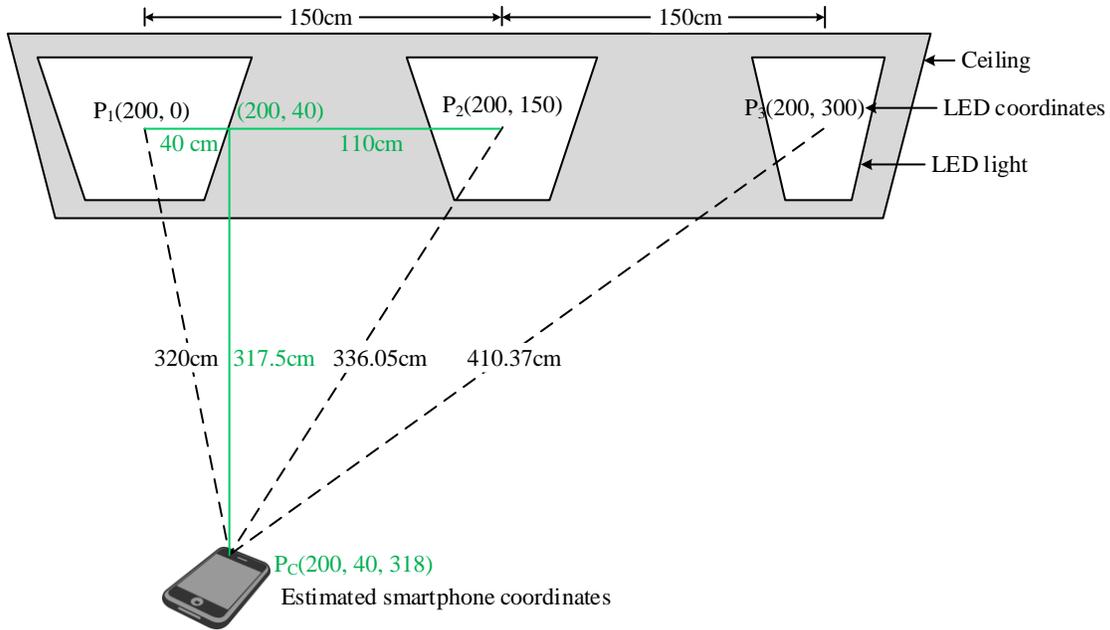

FIGURE 15: Lighting server estimates the coordinates of the smartphone.

In Figure 15 shows an example of the final stage of a server-side process for estimating the smartphone coordinates. Three LED lights are imaged within the FOV of the camera. The distance between each LED light is equal and in our tests this value is 150 cm. The LED light coordinates are $P_1(200, 0)$, $P_2(200, 150)$, and $P_3(200, 300)$, all in cm. Here, the x-coordinates are same for these three LED lights but the y-coordinates are all different. We chose these coordinates to simplify this example. The z-coordinate is equal for all the LED lights, so we ignore it in our calculations.

Let us consider a smartphone placed between $P_1(200, 0)$ and $P_2(200, 150)$ and far away from $P_3(200, 300)$. More precisely, this smartphone is closer to $P_1(200, 0)$ than $P_2(200, 150)$. The distances from the camera to $P_1(200, 0)$, $P_2(200, 150)$, and $P_3(200, 300)$ are 320 cm, 336.05 cm, 410.37 cm, respectively, which

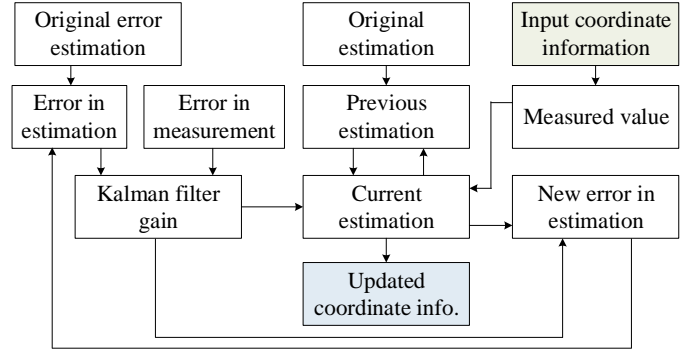

FIGURE 16: Flow diagram of Kalman filter.

are measured by calculating the image sizes on the image sensor. The relative distances from these three LED lights show that the smartphone y coordinate is 40 cm away from $P_1(200, 0)$ and is 110 cm away from $P_2(200, 150)$. In this example we consider that $P_1(200, 0)$ is located at the origin and the y coordinate of the smartphone is calculated with respect to it. Here, 40 cm is the y coordinate of the smartphone. The z-coordinate of the smartphone camera can be measured with the Pythagorean theorem. Therefore, the z-coordinate of smartphone can be calculated as 317.5. Finally, estimated coordinate is $P_C(200, 40, 318)$.

*5.5. Estimate the next location of the smartphone.* After finalizing the smartphone coordinates, the lighting server sends coordinate information to the smartphone, updates this information and stores it information for future use. The location of the smartphone is always changing. While the server estimates the smartphone position, the smartphone may have moved. It is required to run another server-side algorithm in parallel to estimate the velocity, acceleration and next possible position of the smartphone. We use a Kalman filter to track the next position of the smartphone. This filter depends on the present input measurement instead of previous information (e.g., velocity, acceleration) from the smartphone [39].

Kalman filter is a recursive estimator and linear filter mostly used to approximate errors in navigation applications with minimum variance estimate in a least squares sense under noise processes. Kalman filter gain, current estimation, and new error in the estimation are three important calculations in Figure 16. The Kalman filter gain places the special importance on the error in the estimate and the error in the measurement. On the other hand, the current estimation depends on the previous estimation and the present measured value. The relative importance between previous estimate and present measured value is also fixed by the Kalman filter gain. Furthermore, Kalman filter gain and current estimation is needed to know the new error in the estimate which is passed onto error in future the estimate. The preliminary estimated location of the smartphone can describe as

$$X_{k_p} = BX_{k-1} + w_k \tag{31}$$

where $X_{k-1}$ is the initial location of the smartphone, $B$ is the state (or adaptaion) matix, and $w_k$ is noise added to the initial location.

The measurements and state vectors are weighted by their respective processes' covariance matrices. The process covariance matrix (or error in the position estimation) can be represented as

$$P_{k_p} = BP_{k-1}B^T + Q_k \tag{32}$$

where the initial process covariance matrix is $P_{k-1}$ and $Q_k$ is noise.

The filter de-weights the measured value during large variance and low gain in comparison to the state estimate. This situation leads the filter to prioritize the prediction state rather than measurements. In different circumstances, the measured value is weighted more over the predicted value due to the small variance and high gain. The gain of the Kalman filter is known as the Kalman gain which depends on the error in the estimate and error in the measurement. Kalman gain, $K_g$ is the ratio of the error in estimate to the total error in both the estimate and measurement,

$$K_g = \frac{P_{k_p} H^T}{H P_{k_p} H^T + R} \tag{33}$$

where $R$ is the observation or measured error and $H$ is transformation matrix, which converts the covariance matrix into Kalman filter gain matrix. The value of Kalman gain lies between 0 and 1 (i.e., $0 \leq K_g \leq 1$). If $K_g$ is near to 1, it means error in the measurement is nearly 0. In this estimation, the estimates are unstable (large error in the estimate) and measurement are accurate.

Table 1: Physical parameter for the simulated scenarios.

| Camera parameters | |
|---|---|
| FOV | 120° |
| Focal length | 5 mm |
| Pixel size | 1 μm |
| Image size | 640x320 pixels |
| Pixel edge length | 7.1e-3 mm |
| Frame rate | 30 fps |
| Lens aperture | 4 |
| **Parameters for transmitter** | |
| LED diameter | 170 mm |
| LED area | 22,700 mm$^2$ |
| Half power emission | 1500 mW |
| Radiation semi-angle | 20° |
| Size of LED panel | 10x10 cm$^2$ |
| Modulation method | OOK |

The error in location estimate will decrease when the value of $K_g$ is close to 0. Therefore, the difference between estimation and actual is narrowed down. The expression for current estimation can be written as

$$X_k = X_{k_p} + K_g \left[ Y_k - H X_{k_p} \right] \tag{34}$$

where $X_k$ is the present estimate, $X_{k_p}$ is the previous estimate, and $Y_k$ is the measured smartphone coordinate.

Similarly, if the Kalman filter gain is large, then the present error in the estimate is small. The new predicted state can be defined as follows

$$P_k = [I - H K_g] P_{k_p} \tag{35}$$

*5.6. Postponed signal propagation.* The smartphone location measurement and signal propagation stops if the smartphone user leave the room. To recognize this situation, the lighting server broadcasts a message several times and waits for a reply. In the event of no reply, the server stops sending data and stores the position information.

## 6. Simulation Result

To evalute the performance of our proposed scheme, we used a smartphone in 1600 sqft. indoor environment. The test instrument specification are provided in Table 1. The simulation result will varying with the variation of the camera and luminaire parameters.

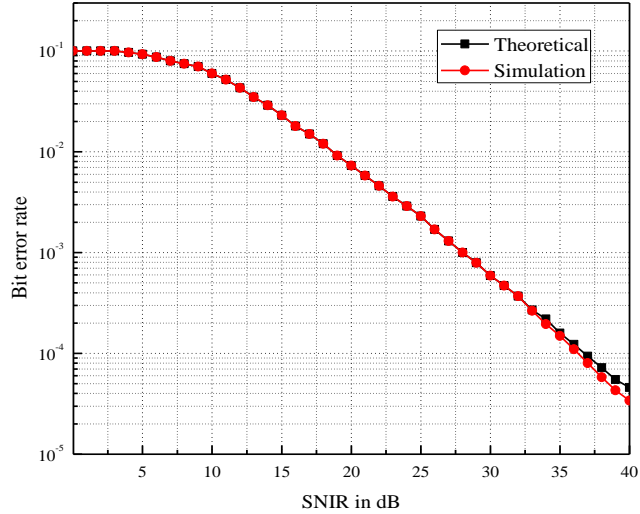

FIGURE 17: Simulation BER vs. theoretical BER with SNIR for OOK signaling.

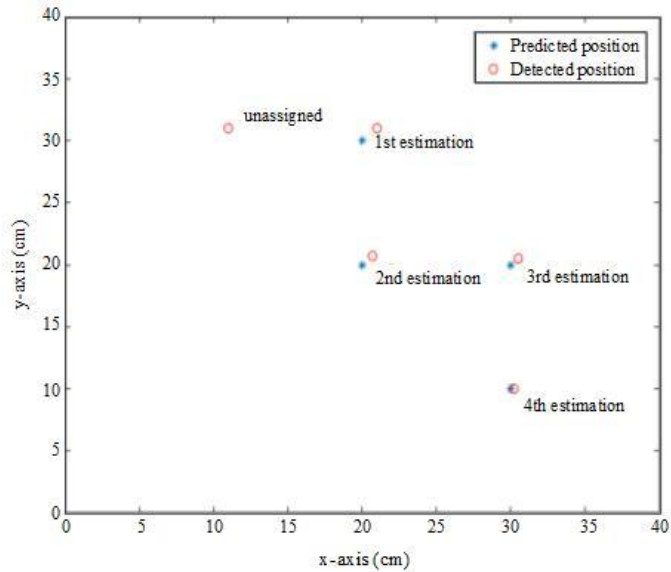

FIGURE 18: Predicting smartphone location using Kalman filter with the detected locations.

Figure 17 shows a graphical representation between BER vs SNIR for theoretical and simulation results. Both curves are almost merged together because we ignore the effect of channel noise in our simulation. It explains that BER is increased with the decrease for SNIR of OOK signaling.

In Figure 18, the simulation result shows that initially the location of the smartphone is not identified and is mentioned as unassigned. Estimation accuracy was not good enough for the 1st estimation compared with the 4th estimation. This estimation process is improved sequentially after a few steps. Concurrently, the distance between each estimation is kept between 9 to 10 cm.

The possibility of changing coordinates in the z-direction is negligible. Therefore, we only have to calculate the x- and y-coordinates of the smartphone. In Figure 19, the green line shows the mean value of the smartphone location and red line is the estimated value. The location estimation using Kalman filter for x axis is plotted in Figure 19(a) and y axis of smartphone are plotted in Figure 19(b). Figure 19 states there is a deviation of location estimation from the mean value of the location. We consider 1 Hz sampling rate and run time is 50 sec. Overall 50 samples were considered for simulation.

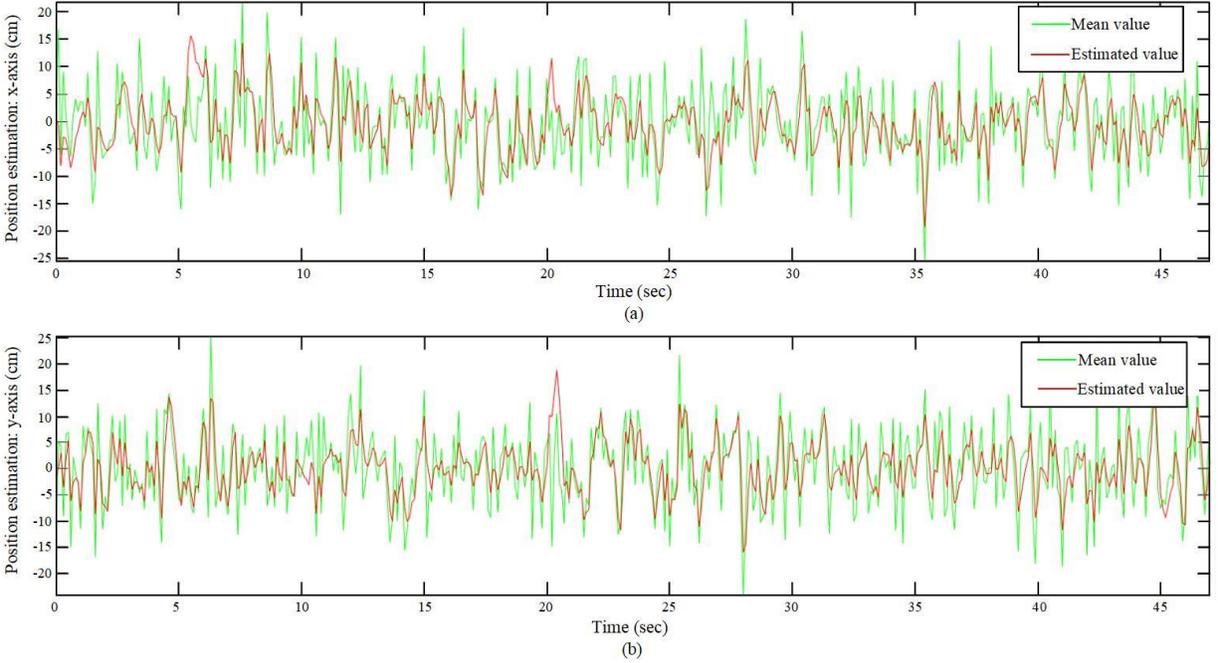

FIGURE 19: Tracking the mean value of the smartphone location using Kalman filter at the direction of (a) x-axis and (b) y-axis.

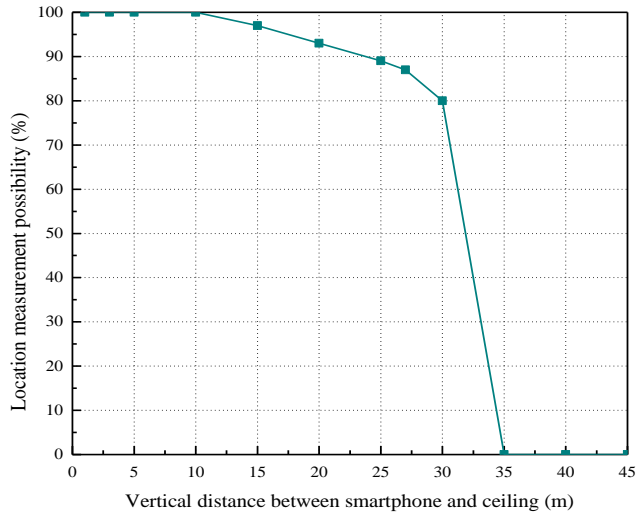

FIGURE 20: Comparison between indoor vertical distance between smartphone and ceiling vs localization measurement possibility.

Distance measurement using OCC depends on the size of the projected image of the LEDs on the IS. With increasing the distance, the projected image on the IS occupies less area rather than a shorter distance. Therefore, the possibility of smartphone localization is shrunk if the vertical distance between the smartphone and the LED lights at the ceiling is increased. In Figure 20, when the vertical distance from the camera to ceiling is remaining within 10 m, the occupying image area is greater or equal 4-pixel area. After 10 to 35 m, localization possibility is reduced due to decreasing occupying pixel area ( $4 > \eta_i \geq 1$ ). Theoretically, it is required to occupy at least unit pixel area of an image sensor. However, it is difficult to ensure that the projected image merges with edge-by-edge of a pixel. Therefore, after 35 m, localization possibility is zero due to occupying pixel area remained $\eta_i < 1$. We consider a fixed transmitter size and in that case, its image occupies less than unit pixel area after 35 m. If we change the transmitter size, then the distance measurement performance will be changed.

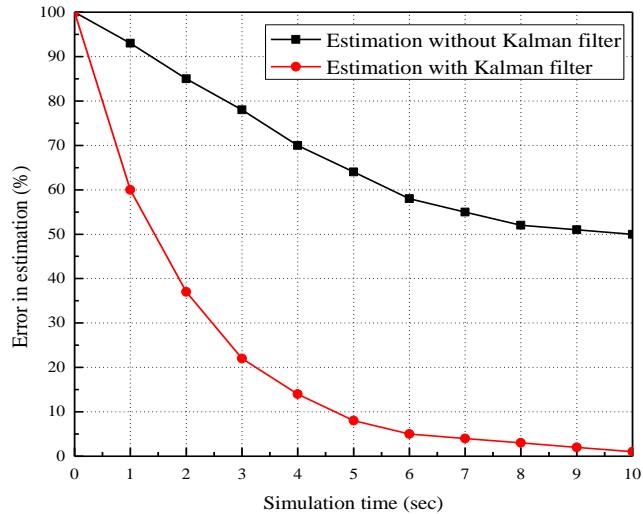

FIGURE 21: Simulation time vs error in estimation on the basis of using Kalman filter.

Localization estimation error occurs due to change of smartphone position frequency. We test our algorithm both with Kalman filter and without Kalman filter. Then we generate a plot for comparing the significance between them in Figure 21. A significant deviation on performance was found from the figure. At the initial stage of measurement, both show the same percentage of errors in estimation. Whereas, in both cases, estimation errors are exponentially decreasing with simulation runtime. At 10 sec period, estimation error is near about zero for Kalman filter based estimation. Meanwhile, at the same time, other estimation (i.e., without Kalman filter) shows 50 % error in estimation.

## 7. Conclusion

In this paper, we proposed a smartphone localization system for an indoor environment. Using OCC for smartphone localization is a novel idea. We also use photogrammetry technique along with OCC. The localization resolution for the smartphone is kept within 10 cm. The proposed system relies upon a central processing lighting server for positioning calculations. Signaling from LED light fixtures and localization of smartphone are kept within certain indoor environment. Therefore, this localization scheme is more secure. Additionally, chance of error in the position estimation is more for the system where the implication of Kalman filter is ignored. We included Kalman filter to track the next possible location of the smartphone. Thus, the proposed scheme is more accurate than existing localization scheme. These lighting fixtures not only useful for localization but also for illumination for the user. In future work, we will test and evaluate the performance in different environmental scenarios e.g., escalator, staircase. We will consider variation in height between smartphone and ceiling light fixtures. Meanwhile, we are also trying to optimize the position identification resolution without using Kalman filter to make the system simpler.

## Conflicts of Interest

The authors declare that they have no conflicts of interest.

## Acknowledgments

This research was supported by the MSIT (Ministry of Science and ICT), Korea, under the Global IT Talent support program (IITP-2017-0-01806) supervised by the IITP (Institute for Information and Communication Technology Promotion).